\documentclass{jpsj2}
\title{Energy Dissipation Burst on the Traffic Congestion }

\author{Kaito \textsc{Umemura}$^{1}$\thanks{E-mail address: umemura@radix.h.kobe-u.ac.jp} and Kuniyoshi \textsc{Ebina}$^{2,3}$}

\inst{$^{1}$Graduate School of Cross Cultural Studies and Human Science, Kobe University, Kobe, 657-8501 \\
$^{2}$Faculty of Human Development, Kobe University, Kobe, 657-8501 \\
$^{3}$Graduate School of Science and Technology, Kobe University, Kobe, 657-8501}

\abst{We introduce an energy dissipation model for traffic flow based on the optimal velocity model (OV model).
In this model, vehicles are defined as moving under the rule of the OV model, 
and energy dissipation rate is defined as the product of the velocity of a vehicle and resistant force which works to it.
According to the results of numerical simulation on the periodic boundary condition, 
the energy dissipation depends on traffic conditions, such as congestion (traffic jam), 
and therefore depends on parameters and initial conditions of the system.
The results of simulation indicate that, 
although the flux of vehicular transportation is not so different between at a state of free flow and that of congestion, 
the energy dissipation reveals a burst at a state of congestion. 
Additionally, how the burst behaves depends on how congestion appears. 
}

\kword{traffic flow, optimal velocity model, energy dissipation, congestion, steady state flow}

\begin{document}
\maketitle
\pagebreak

\section{Introduction} 
A lot of researches have been done on traffic flow model since 1950s when traffic systems were developing in the real world. 
From the viewpoint of statistical physics, lots of traffic flow models were proposed\cite{1,2}$^{)}$. 
The optimal velocity model (OV model)\cite{3,4}$^{)}$ and the cellular automata model (CA model)\cite{5,11}$^{)}$ are examples of micro models, 
and fluid model is an example of macro models\cite{6,9}$^{)}$. 

In the previous researches of traffic flow, 
one of the main questions is how traffic congestion (traffic jam) occurs and whether it is stable or not. 
In the research of OV model for example, the method of linearizing analysis is applied 
and the conclusion is that congestion is understood as some kind of phase transition of a system 
and its stability depends on parameters as well as initial and boundary conditions. 
It means that, at least in the conventional OV model, congestion is intrinsic phenomenon of a system 
and can appears without external forcing like bottleneck of road. 

In the previous studies, 
many works were done on vehicle's flux of spatial transportation and stability of state, but almost none on energy dissipation. 
Originally, research of traffic flow has much relation to social issues like energy problems, 
so we think it is important to consider the energy dissipation in modelling the traffic flow. 

In the viewpoint of engineering, 
the researches of one vehicle, including measurement of fuel efficiency, have been done by many groups in automobile industry. 
But there are few discussions of fuel efficiency of the total system of vehicles or their energy dissipation 
because the whole system is too large to measure in the real world. 
So it is important to estimate what happens in the real traffic 
by modelling the whole traffic systems and calculating the energy dissipation of them.  

We would like to understand physical feature of energy dissipation of traffic flow 
but it is not obvious from equation of motion of traffic flow model. 
So we have to make a new modelling of energy dissipation combining with a former model of vehicle's motion. 
We use the OV model to combine with our energy dissipation model. 
The OV model is simple but describes well the appearance of congestion cluster in the system, 
thus it suits our energy dissipation model. 
Although it has been pointed out that the OV model has some problems by comparison with empirical data, 
we use this model in this paper \cite{7}$^{)}$.  

In the following, we first describe the OV model on which we base (\S 2.1) and then propose an energy dissipation model (\S 2.2). 
Next, we show a results of numerical simulation (\S 3) and do some discussions about the results (\S 4). 
Finally we give summary (\S 5).

\section{Models}
\label{}

\subsection{Optimal velocity model}
The optimal velocity model (OV model) was proposed by Bando et al. in 1995 
and it has been used and modified by many researchers in various ways until today\cite{3,4,7}$^{)}$. 
The OV model is one of micro models and it defines the dynamical equation of motion for each vehicle in one dimensional space. 
The equation of motion is given as 
\begin{gather}
\frac{d^2}{d t^2} x_n(t) = a\left[V(\Delta x_n(t)) - \frac{d}{dt} x_n(t)  \right], 
\end{gather}
where 
\begin{gather}
\Delta x_n(t) = x_{n+1}(t) - x_n(t)
\end{gather}
for each vehicle number $n$ $(n = 1,2,...,N)$.
$N$ is the total number of vehicles, 
$x_n$ is the coordinate of the $n$th vehicle and is a function of time $t$, $\Delta x_n$ is its forward distance to the preceding $(n+1)$th vehicle, 
and $a$ is a parameter called sensitivity which represents driver's response speed. 
We assume that $\Delta x_n$ should be positive. 
The most important feature of this model is assuming the ``optimal velocity" function $V(\Delta x_n)$.
The OV function is a function of the forward distance $\Delta x_n$ of vehicle number $n$, and having the properties: 
(i) a monotonically increasing function, 
(ii) $|V(\Delta x_n)|$ has lower and upper limit. 
The upper limit of the OV function corresponds to $v_{\text{max}} = V(\Delta x_n \rightarrow \infty)$.
We adopt a functional form of $V(\Delta x_n)$ as  
\begin{gather}
V(\Delta x_n) = \frac{v_{\text{max}}}{2} \left[ \tanh \left(\frac{\Delta x_n - c}{w} \right) + \tanh \left(\frac{c - d}{w} \right) \right],
\end{gather}
which satisfies the properties (i) and (ii).
Parameters $c$, $d$, and $w$ control the shape of OV function: 
$c$ corresponds to the $\Delta x$ at inflection point of OV function 
representing the forward distance in which the sign of increasing rate of vehicle's optimal velocity changes; 
$d$ represents the forward distance where optimal velocity of vehicle becomes zero; 
and $w$ determines the slope of OV function. 
They all have dimensions of length. 
Fig. 1 shows the shape of $V(\Delta x)$.

\begin{figure}[h]
\begin{center}
\includegraphics[scale=0.7,clip,angle=-90]{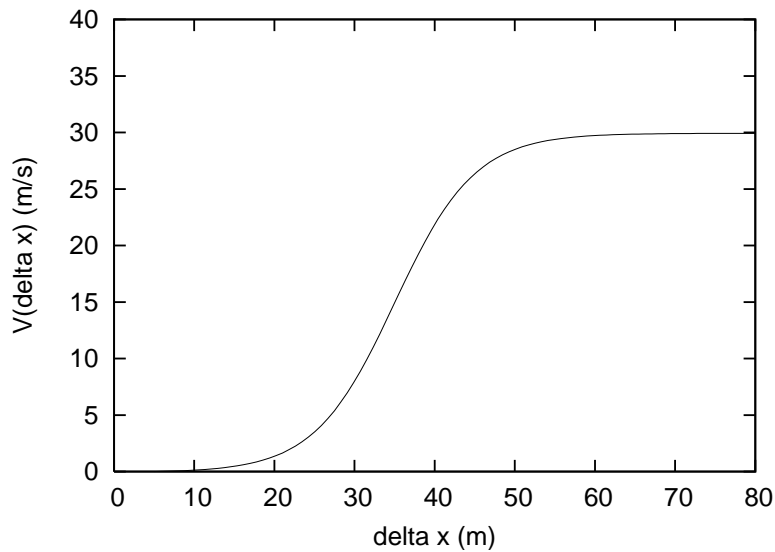}
\end{center}
\caption{An example of OV function given in eq. (3) with $v_{\text{max}}=30$ m/s, $c=35$ m, $d=4$ m, $w=10$ m. 
It has a sigmoid shape having one inflection point at $\Delta x = c$. }
\end{figure}

\subsection{Modelling the energy dissipation}
Now, we introduce our energy dissipation model.
Vehicles move in resistant forces like air drag, friction between road and tires, and so on. 
Conversely speaking, vehicles do work to external environment. 
These works will finally be converted to heat and dissipate in the air. 
That brings transport of energy from vehicles to the air. 
Therefore we define energy dissipation rate of each vehicle $j_{q}$ 
as the product of vehicle's velocity $v$ and the total resistant force $F_r$ which work to each vehicle: 
\begin{gather}
j_{q} = F_{r} v .
\end{gather}
In this model, thermal efficiency of engine is not taken into account. 
We shall examine the validity of this approach later in \S 5. 
Then we have to model the resistant force $F_r$ working to each vehicle to consider energy dissipation 
as a function of the state of a vehicle. 
We simply assume that $F_r$ consists of three parts, defined as follow: 
\begin{gather}
\begin{split}
F_{r} &= F_a + F_f + F_b \\
      &= (\alpha v + \beta v^2) + f + F_b.
\end{split}
\end{gather} 
$F_a(= \alpha v + \beta v^2)$ is the air drag, which is a function of the velocity of vehicle $v$.
$F_f$ contains other frictions working to vehicle, which we assume being constant $f$.
$F_b$ is the braking force controlled by the driver, which only appears when vehicle reduces its speed. 

The problem here is how to define the functional form of $F_{b}$ for eq. (5) where we have a freedom of choice. 
To solve this problem, we consider the Newton's equation of motion of a vehicle. 
We define $F_e$ as the force created by engine to move a vehicle ahead, 
then Newton's equation of motion of a vehicle may be written as 
\begin{gather}
M \frac{d v}{d t} = F_e - F_r
\end{gather} 
where signs of $F_r$ is defined as positive when its direction is opposite to vehicle's moving direction, 
while that of $F_e$ being positive when its direction is equal to vehicle's moving direction.
$M$ is the mass of the vehicle considered.
Considering the OV model, eq. (1), and the frictional force, eq. (5), the following equation is derived: 
\begin{gather}
Ma[V(\Delta x) - v] = F_e - F_b - (\alpha v + \beta v^2) - f.
\end{gather} 
When a vehicle is accelerating or moving with constant speed, 
the value of $F_{b}$ must be zero and the functional form of $F_{e}$ is determined from eq. (7). 
But $F_{b}$ is not zero when a vehicle is decelerating, and the functional form of $F_{e} - F_{b}$ is determined as 
\begin{gather}
F_{e} - F_{b} = (\alpha v + \beta v^2) + f + Ma[V(\Delta x) - v]. 
\end{gather} 
We have to model $F_{e}$ and $F_{b}$ separately to obtain a functional form of $F_{b}$. 
Here, we set two types of assumption to separate them as: 
\begin{align}
F_{b} &= - Ma[V(\Delta x) - v] \\
F_{e} &= (\alpha v + \beta v^2) + f
\end{align}
for model type 1, and
\begin{align}
F_{b} &= - (\alpha v + \beta v^2) - Ma[V(\Delta x) - v] \\
F_{e} &= f
\end{align}
for model type 2.
Type 1 means that when a vehicle is reducing its speed, 
the braking force $F_{b}$ is used for decelerating and the driving force $F_{e}$ is for air drag and frictions.
Type 2 means that air drag is also used for decelerating and $F_{e}$ is just for frictions. 
So to speak, type 1 and 2 represent possible maximum and minimum power of engine respectively when a vehicle is decelerating. 
Finally the total resistant force of $n$th vehicle $F_{r}^{(n)}$ is determined in three cases as follow 
depending on the sign of vehicle's acceleration and the type of $F_{b}$, 
\begin{align}
\text{(when accelerating or moving with } & \text{constant speed)} \notag \\
F_{r}^{(n)} =& F_a + F_f \notag \\
            =& (\alpha v_n + \beta v_n^2) + f_{n} \\
\text{(when decelerating by $F_{b}$ of type 1)} \notag \\
F_{r}^{(n)} =& F_a + F_f + F_b \notag \\
            =& (\alpha v_n + \beta v_n^2) + f_{n} - M_{n}a[V(\Delta x_n) - v_n] \\
\text{(when decelerating by $F_{b}$ of type 2)} \notag \\
F_{r}^{(n)} =& F_a + F_f + F_b \notag \\
            =& f_{n} - M_{n}a[V(\Delta x_n) - v_n]. 
\end{align} 
Here, $M_{n}$, $v_{n}$ and $f_{n}$ are the mass, the velocity and the constant friction of the $n$th vehicle respectively.
According to eq. (11), $F_{b}$ of type 2 can be negative when the vehicle decelerates 
if $(\alpha v + \beta v^2) > - Ma[V(\Delta x) - v]$. 
Therefore we supplement to the model one additional rule that $F_{b}$ is replaced with zero if $F_{b} < 0$. 

As a consequence of above formulation, 
the energy dissipation per unit time of $n$th vehicle $j_{q}^{(n)}$ and of the entire system $J_{q}$ is given as
\begin{gather}
j_{q}^{(n)} = F_{r}^{(n)} v_n \\
J_q = \sum_n j_{q}^{(n)} = \sum_n (F_{r}^{(n)} v_n) .
\end{gather}
We use eqs. (16) and (17) for calculation.

\section{Numerical Simulation and Results}
\label{}

\subsection{Algorithms}
\subsubsection{Models}
We numerically solve the differential equation (1), 
and calculate the energy dissipation rate for each vehicle and the total system using eqs. (16) and (17).
We then integrate numerically the energy dissipation rate with respect to time 
for getting the total dissipated energy $E$ during all the simulation time $T$.
Then the time average of $J_{q}$ is obtained as 
\begin{align}
\text{(time average of $J_{q}$)} = \langle J_{q} \rangle &= \frac{E}{T} \notag \\
                      &= \frac{\int J_q dt}{T} \notag \\
                      &= \frac{\int \sum_n (F_{r}^{(n)} v_n) dt}{T}, 
\end{align}
where $\langle J_{q} \rangle$ has dimension of energy per unit time. 
In addition to the energy dissipation rate of system $\langle J_{q} \rangle$, 
we also need another coefficient 
which has dimension of energy per unit distance to represent energy efficiency of vehicular transportation of system, 
so that we define $e$ which is the time average energy dissipation per distance of one vehicle in system as 
\begin{align}
e &= \frac{E}{X} \notag \\
  &= \frac{\int J_q dt}{X} \notag \\
  &= \frac{\int \sum_n (F_{r}^{(n)} v_n) dt}{X}, 
\end{align} 
where $X$ is sum of achieved distance of all vehicles during all the simulation time $T$. 
The lower $e$ represents more efficient transportation of vehicles. 

In this paper, 
we have main focus on dynamical behavior of $j_{q}^{(n)}$ and the dependence of $\langle J_{q} \rangle$ and $e$ on parameters. 

\subsubsection{Simulation conditions}
We set a periodic boundary condition, 
in which vehicles move around circuit of length $L$ and $(N+1)$th vehicle is identical to the first one.
Furthermore, circuit is regarded as one dimensional and has no passing of vehicles. 
To avoid passing, sensitivity $a$ is set larger than 0.8($1/s$) throughout all simulations. 
We usually set the initial condition of coordinates and velocities of vehicles as 
\begin{gather}
\Delta x_n(0) = x_{n+1}(0) - x_{n}(0) = L/N \\
v_n(0) = V(L/N),
\end{gather}
while we impose fluctuations of initial coordinates in some cases. 

In this paper, we are interested in how macroscopic values of the system depend on $a$ and $\rho$, 
so that other parameters are fixed on constant values.  
Fixed parameters are given simply and semi-realistically: 
(parameters of OV function) $c=35$ m, $d=4$ m, $w=10$ m; 
(parameters of energy dissipation model) $\alpha=0$, $\beta=1.12$ kg m$^{-1}$, 
$M_{n}=1800$ kg, $f_{n} = \mu M_{n} g$, $\mu = 0.01$(friction coefficient), $g = 9.8$ m s$^{-2}$(gravitational acceleration); 
$L=5000$ m (circuit length) \cite{8,10}$^{)}$. 
Using these parameters, we estimate the order of magnitude of each resistant force as 
\begin{align}
F_{a} &= (\alpha v + \beta v^2) \simeq 1000 \text{ N} \\
F_{f} &= f \simeq 180 \text{ N} \\
F_{b} &= - M a \left[V(\Delta x) - v \right] \simeq 18000 \text{ N}, 
\end{align}
where we set $v = 30$ m s$^{-1}$ (maximum speed), $a = 1.0$ s$^{-1}$ and $V(\Delta x) = 20$ m s$^{-1}$. 
The braking force here is that of model type 1.  
This estimation shows that the braking force is much higher than air drag and constant friction. 

\subsection{Results}
At first, we use eq. (14) for resistant force (type 1) and calculate the energy dissipation rate.
We performed a series of simulations of OV model and the calculated energy dissipation rate $j_{q}^{(1)}$ 
for a typical vehicle in a typical situation in Fig. 2. 
Table I is a list of some macroscopic values corresponding to Fig. 2(a)-(d).
The left side of Fig. 2(a)-(d) are plots of positions of all vehicles ($N=120$) on the circuit length $L$ 
with time development $(x_{n}, t)$, 
where the line which goes upper right in each figure is the trajectory of vehicle number one $(x_{1}, t)$. 
The time duration $T$ is fixed as 1000 s throughout the simulations 
because any important event has already happened until the end of this period in almost all cases.  
The right side of Fig. 2(a)-(d) respectively are diagrams of time development of 
$v_{1}$ and $j_{q}^{(1)}$ 
which correspond respectively to the velocity and energy dissipation rate of the vehicle number one in the left figures. 
Fig. 2(a)-(d) are different in the initial positions $x_{n}(0)$ and sensitivity $a$. 
Fig. 2(a) ($a=2.0$ s$^{-1}$) is the result with no fluctuation in initial condition given as eq. (20) and no congestion appears in it.
Fig. 2(b) ($a=1.0$ s$^{-1}$) is the result with $x_{100}(0)$ being deviated by $-20$ m while the homogeneous vehicular separation $L/N \simeq 42$ m and there appears one congestion cluster. 
In Fig. 2(c) ($a=1.0$ s$^{-1}$), $x_{20}(0)$, $x_{60}(0)$ and $x_{100}(0)$ have deviations of $-20$ m and there appears three congestion clusters.
Fig. 2(d) ($a=1.0$ s$^{-1}$) shows an example for which all $x_{n}(0)$ have perturbations given randomly between $-L/2N$ and $+L/2N$, and there appears many congestion clusters.
In the right side of Fig. 2(a), there appears constant velocity and energy dissipation rate with no fluctuations.
On the other hand we see fluctuations of those in the right side of Fig. 2(b)-(d). 
The energy dissipation rate shows ``spike"-like shape and large magnitude when the vehicle diminishes its speed.
We call this effect ``energy dissipation spike" of a vehicle.
These fluctuations are due to the appearance of congestion. 
These results may be summarized that the more congestion clusters appear in the system, 
the shorter the period of fluctuation becomes.

Fig. 3 shows the time average of, 
(a) energy dissipation rate of the system $\langle J_{q} \rangle$, 
(b) that of flux of vehicular transportation $Q$ 
and (c) that of energy efficiency of one vehicle $e$ 
versus the homogeneous vehicle density $\rho$ in circuit, 
with ensemble-averaging of the initial coordinates of vehicles where $x_{n}(0)$ has random fluctuation between $-L/2N$ and $L/2N$.
It shows large increase in energy dissipation rate $\langle J_{q} \rangle$ and energy efficiency $e$, 
as well as two branches of vehicular flux $Q$ in the middle region of $\rho$. 
We call this effect ``energy dissipation burst" of vehicular system. 
This effect represents the situation that higher energy dissipation is inevitable because of appearing of congestion clusters 
while the system maintains its flux of vehicular transportation effectively the same. 
It can be recognized that congestion occurs only in this middle region, 
and there exist low and high density steady state flow, which means density of the system and the velocity of every vehicle are constant, 
in the left and right side of the region. 
Fig. 3(c) shows that, 
in the steady state flow with high density of vehicles, 
the energy dissipation for transporting vehicle is smaller than in the flow with low density. 
However as shown in Fig. 3(b), this region corresponds to much lower transportation rate. 

Fig. 4 is a diagram which shows behavior of $\langle J_{q} \rangle$ and $e$ as a function of $\rho$ and $a$. 
In the figures, it seems that the width of middle region of $\rho$, 
which corresponds to the energy dissipation burst region, is wider when $a$ is smaller.  
And the burst disappears in the region of large $a$, where no congestion occurs. 
The values of energy dissipation rate and energy dissipation per distance are not constant but show ``mountain" like shape inside the burst region. 

Another series of simulation like above have been done using eq. (15) for resistant force. 
The results shows almost the same value and shape, and it is shown in Fig. 3 as an example. 

\begin{figure}[htbp]
\begin{center}
(a)\includegraphics[scale=0.41,clip,angle=-90]{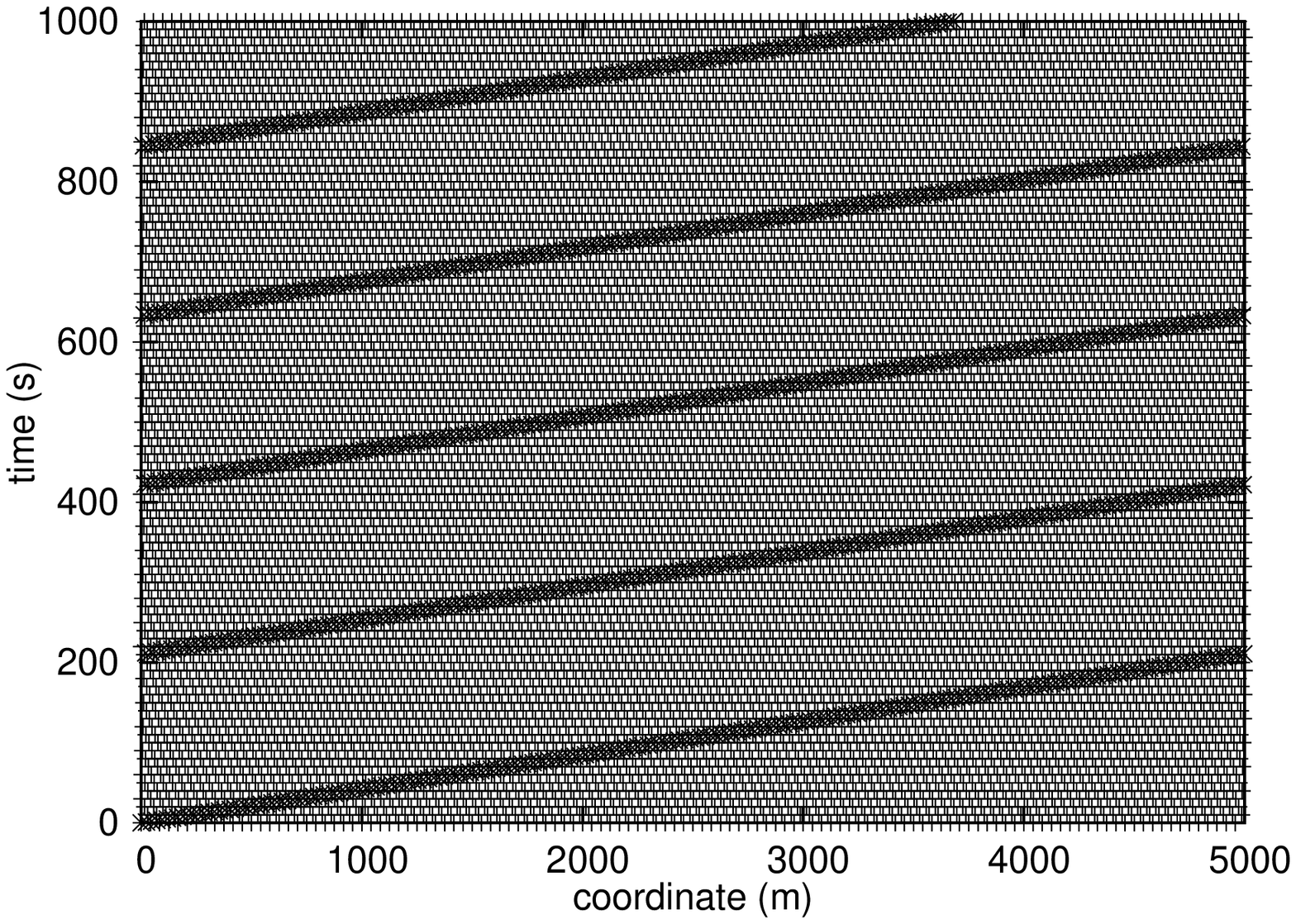}\includegraphics[scale=0.41,clip,angle=-90]{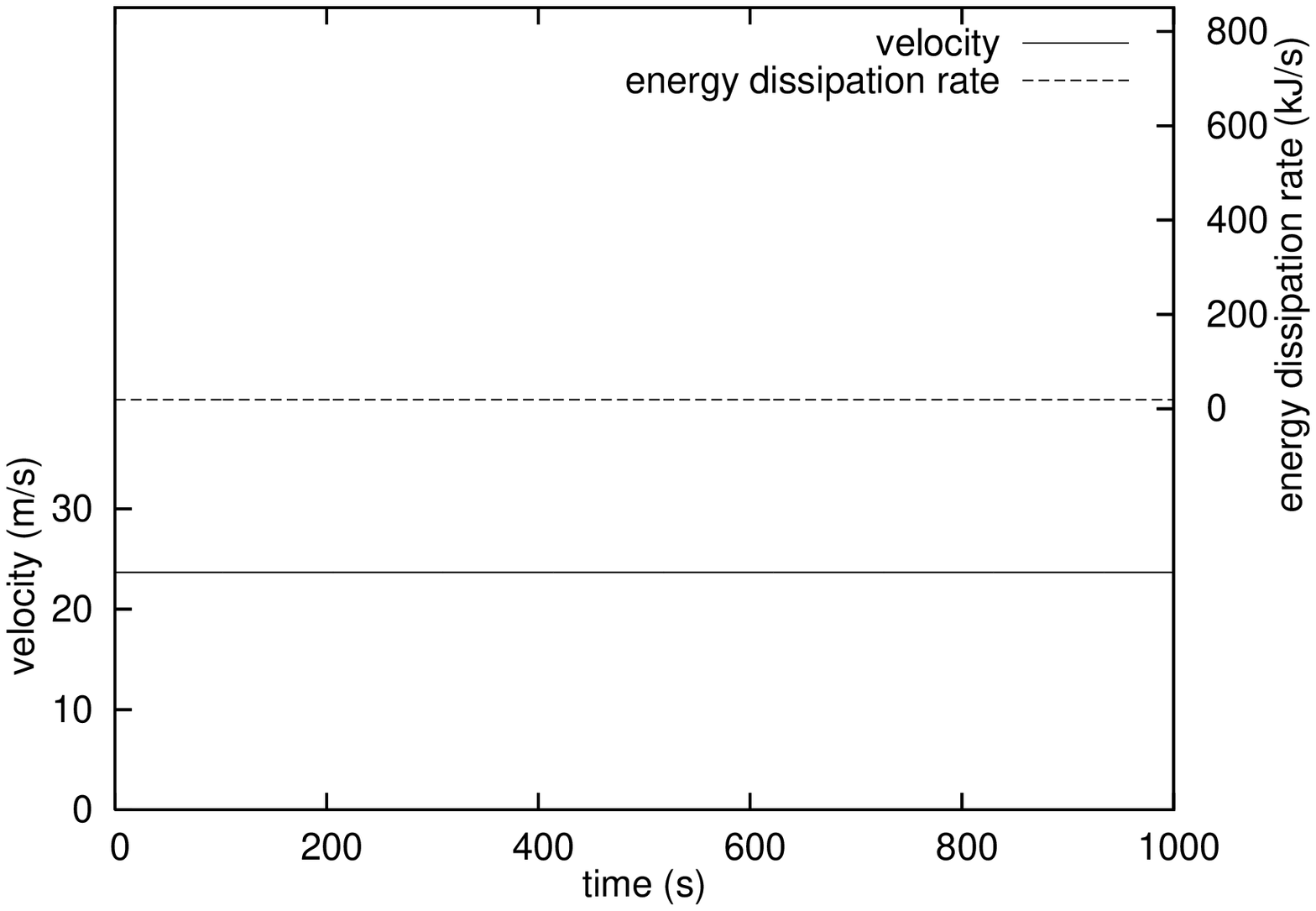}
(b)\includegraphics[scale=0.41,clip,angle=-90]{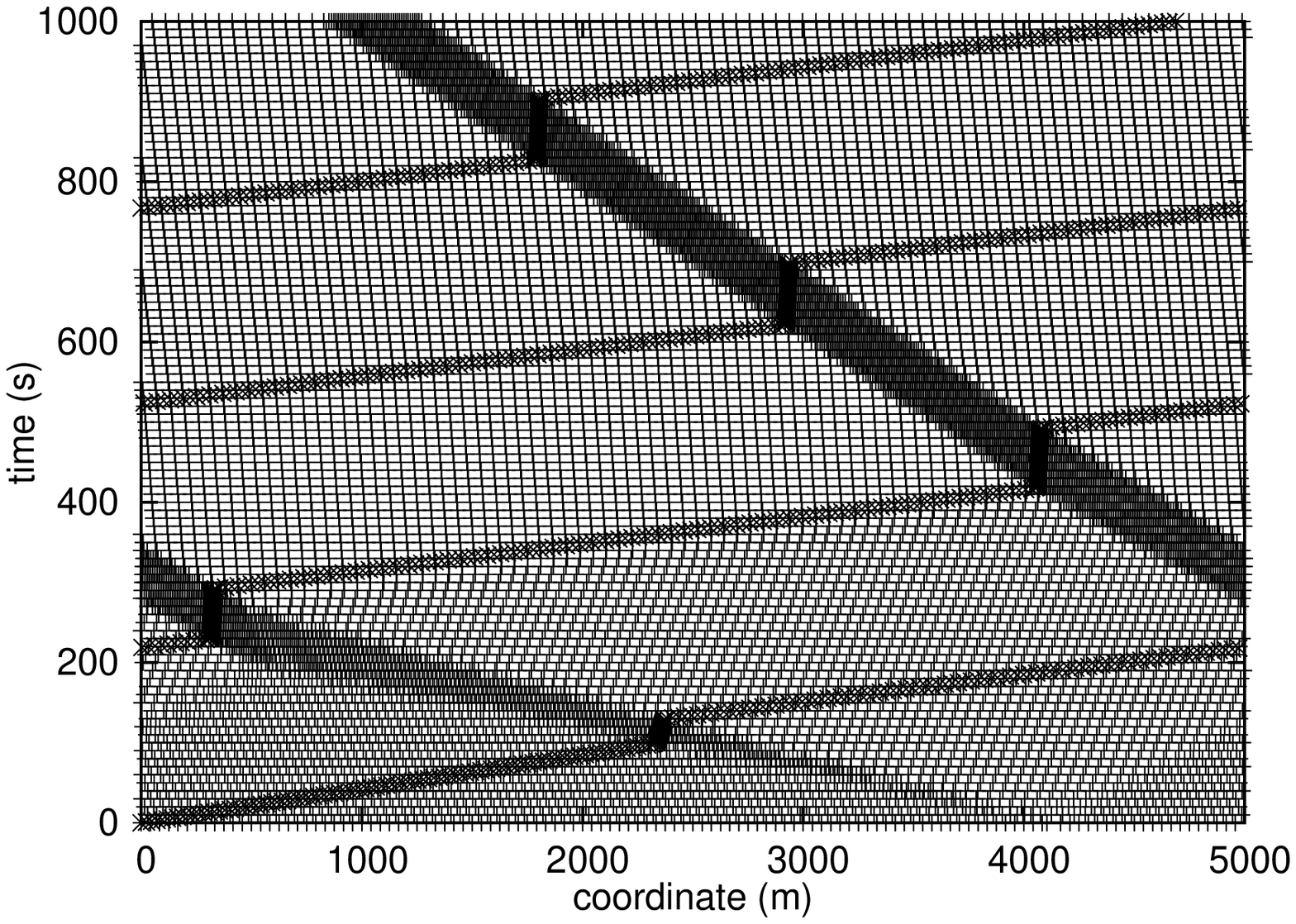}\includegraphics[scale=0.41,clip,angle=-90]{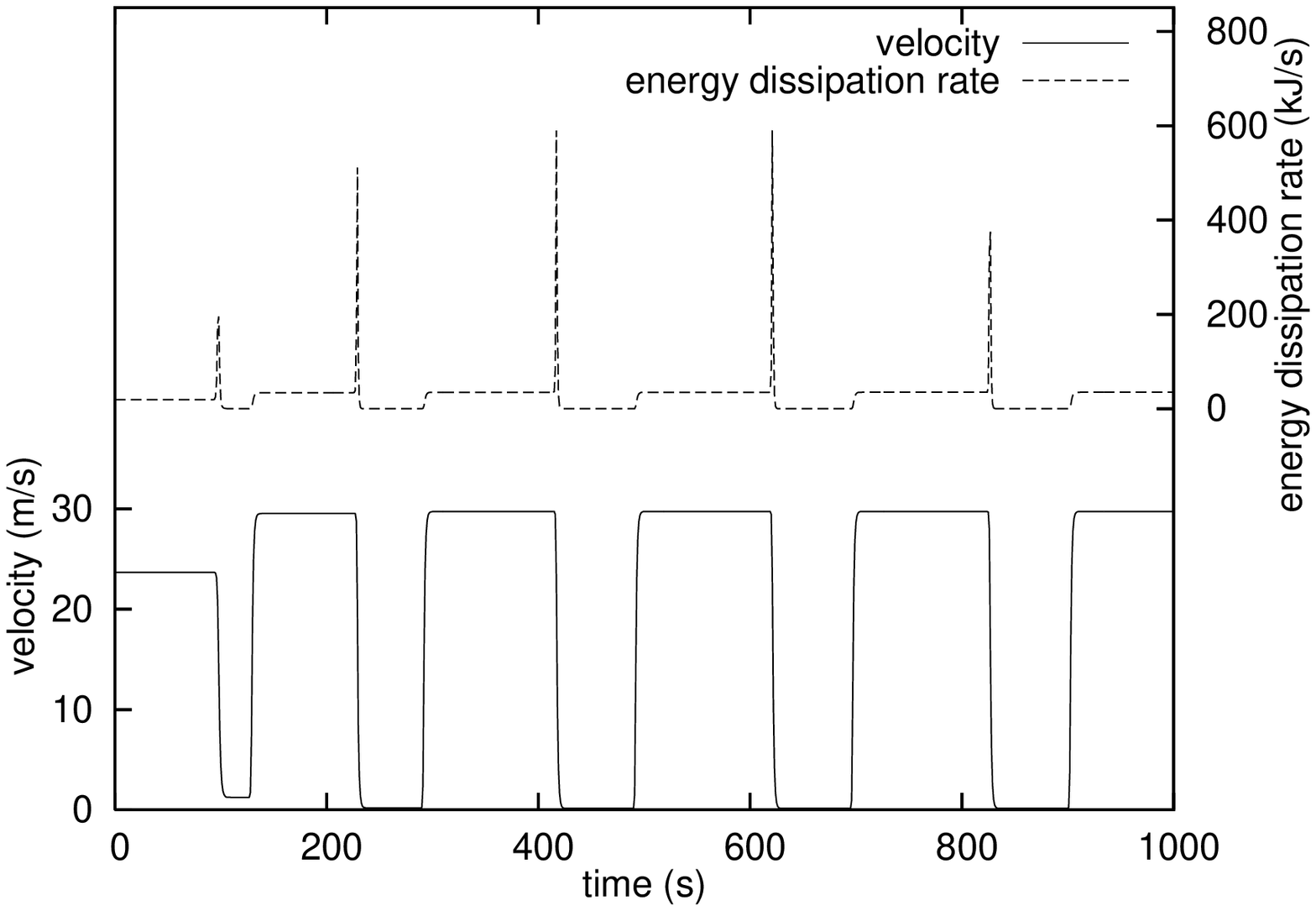}
(c)\includegraphics[scale=0.41,clip,angle=-90]{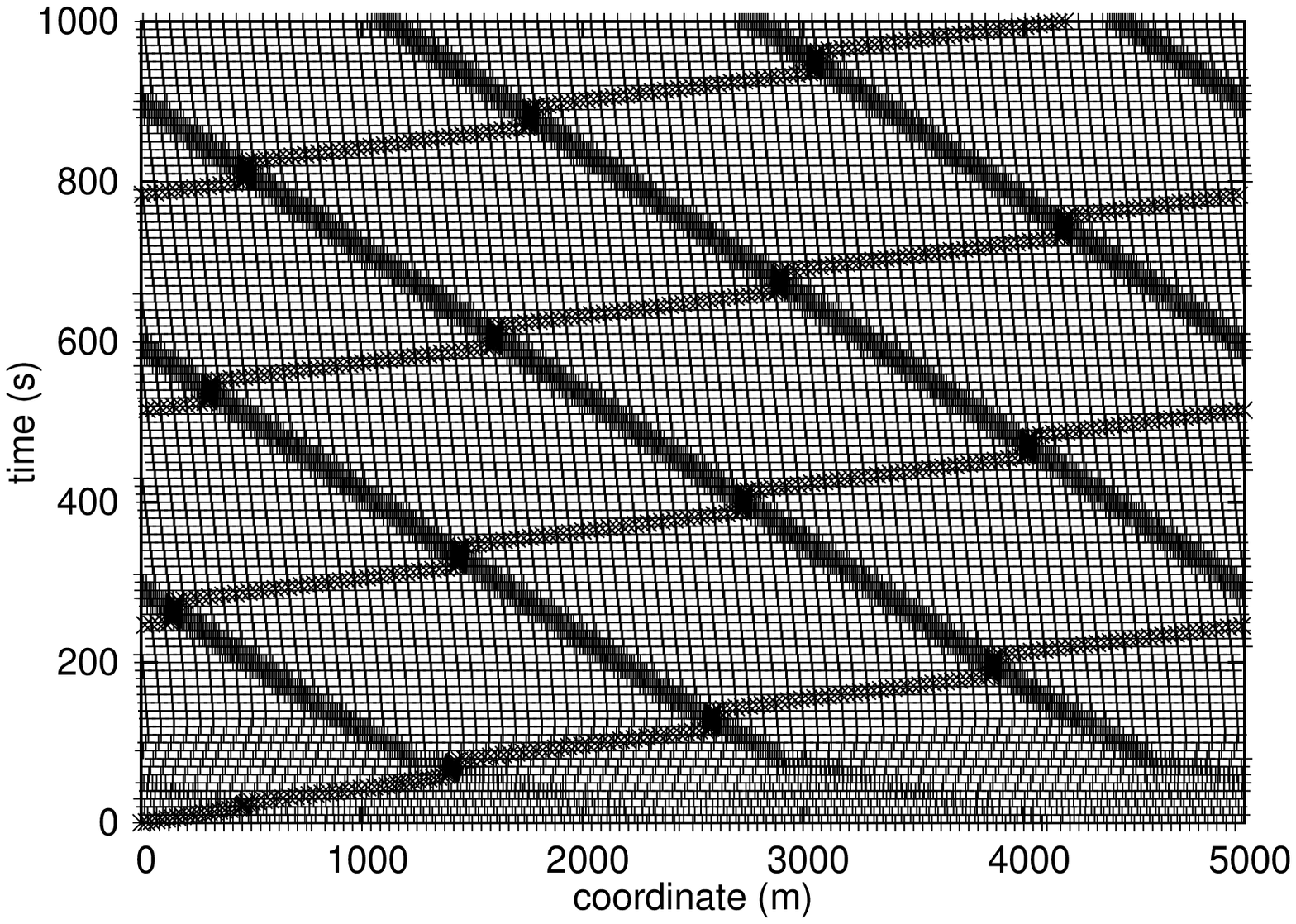}\includegraphics[scale=0.41,clip,angle=-90]{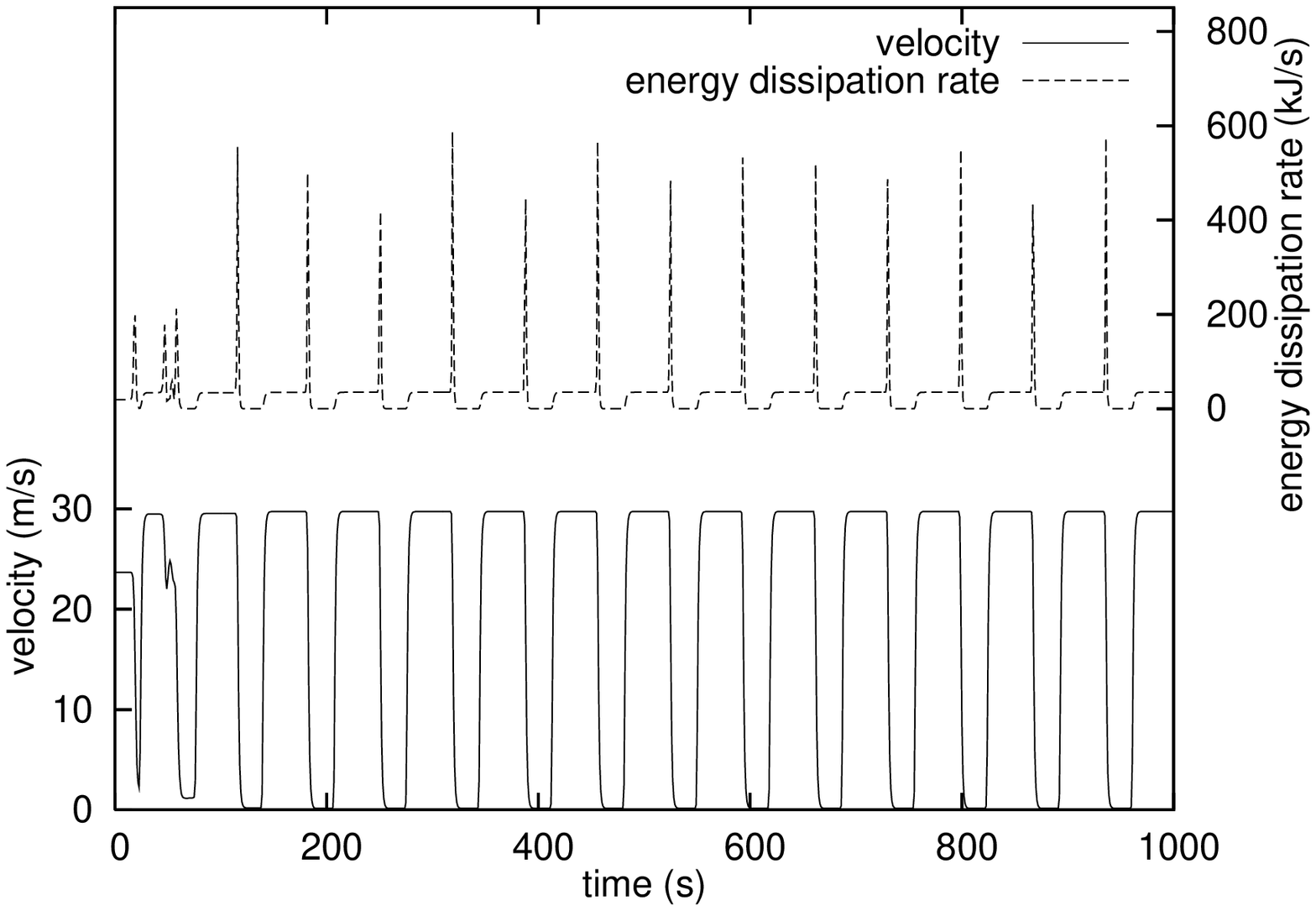}
(d)\includegraphics[scale=0.41,clip,angle=-90]{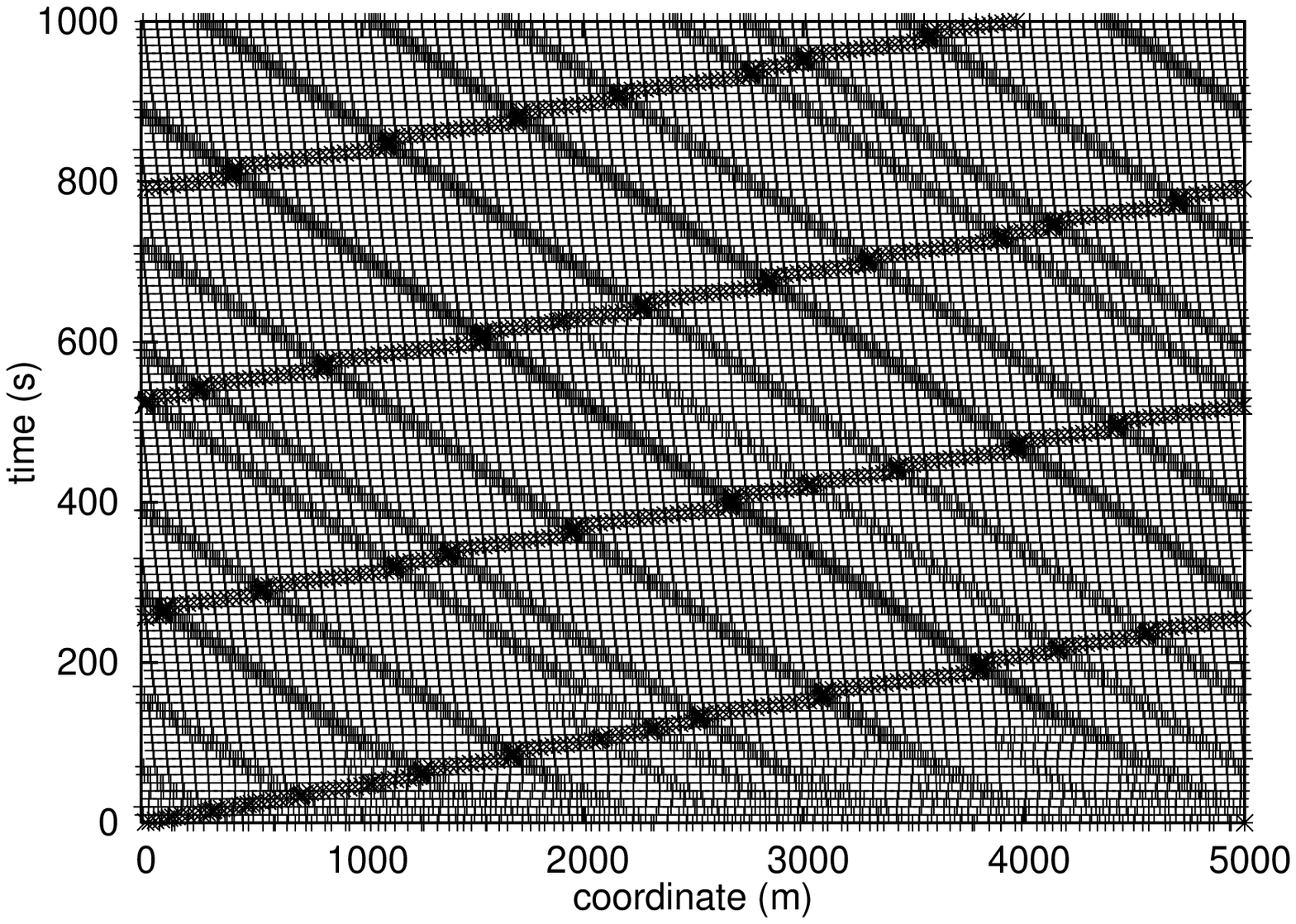}\includegraphics[scale=0.41,clip,angle=-90]{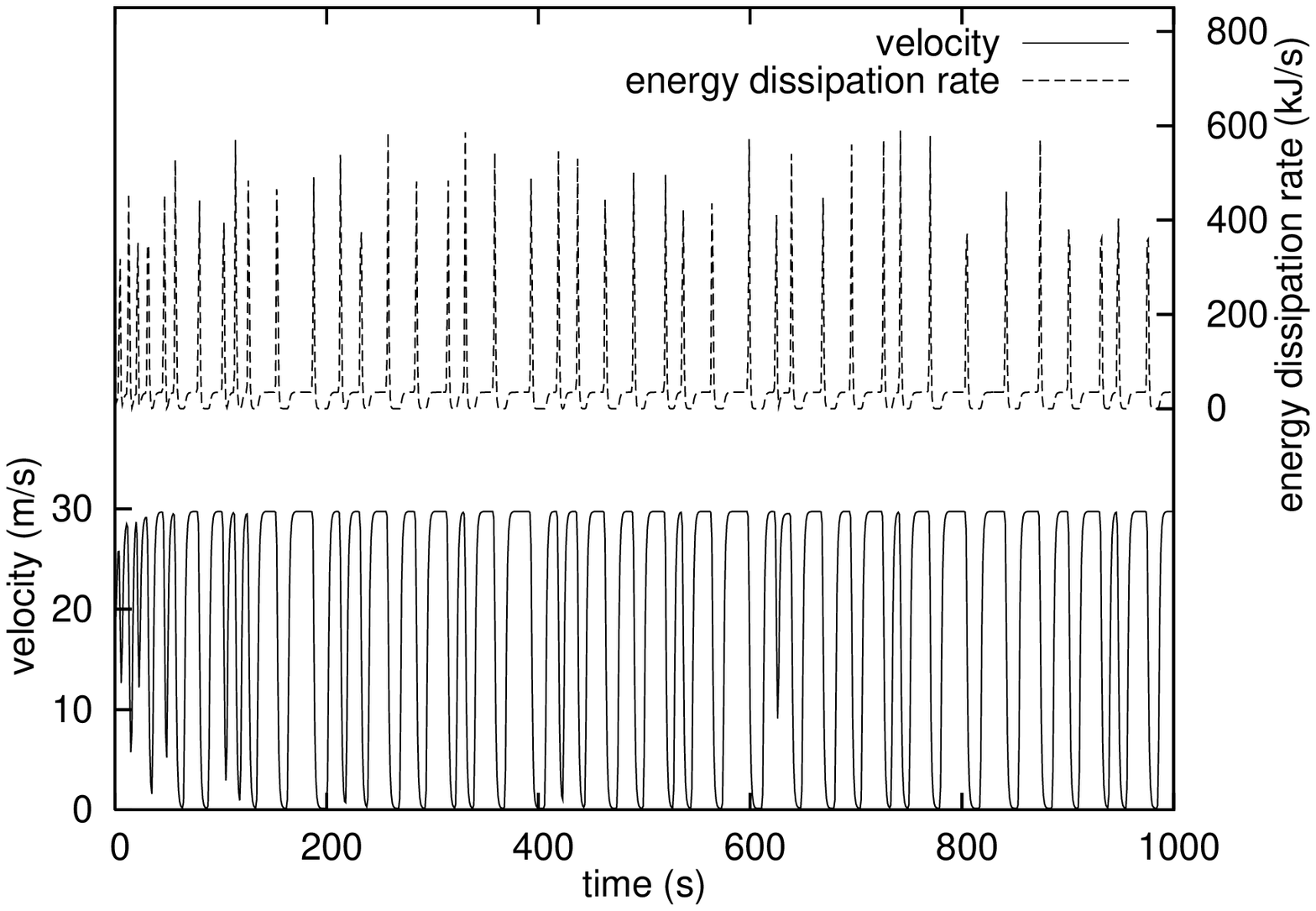}
\end{center}
\caption{(left) time-space diagrams of OV model with the thick lines corresponding to the trajectory of vehicle no. 1; 
(right) time development of the velocity $v_{1}$ and energy dissipation rate $j_{q}^{(1)}$ 
of vehicle no. 1 in each of left figures. 
Fig. (a)-(d) are different in the initial coordinates of vehicles 
and so are the number of congestion clusters in the system, 
all with the same values of the vehicle density $\rho$ and the sensitivity $a$. }
\end{figure}

\begin{table}[tb]
\caption{Some macroscopic values of Fig. 2(a)-(d). 
$\langle j_{q} \rangle$ is the average energy dissipation rate of one typical vehicle, 
$\langle J_{q} \rangle$ is that of total system, 
$Q$ is the average flux of vehicular transportation, 
and $e$ is the value which is derived by $\langle J_{q} \rangle$ (simulation time duration)/(sum of achieved distance of all vehicles), 
an average energy dissipation per distance of a vehicle (the lower the better ).}
\label{t2}
\begin{tabular}{lrrrr}
\hline
value & Fig.2(a) & Fig.2(b) & Fig.2(c)  & Fig.2(d) \\
\cline{2-5}
$\langle j_{q} \rangle$ / kJ s$^{-1}$ & 19.05 & 25.79 & 29.37 & 51.81 \\
$\langle J_{q} \rangle$ / kJ s$^{-1}$ & 2286 & 3095 & 3962 & 6216 \\
$Q$ / vehicles s$^{-1}$ & 0.568 & 0.450 & 0.461 & 0.457 \\
$e$ / kJ m$^{-1}$ & 0.8045 & 1.333 & 1.737 & 2.735 \\ 
\hline
\end{tabular}
\end{table}

\begin{figure}[htbp]
\begin{center}
(a)\includegraphics[scale=0.4,clip,angle=-90]{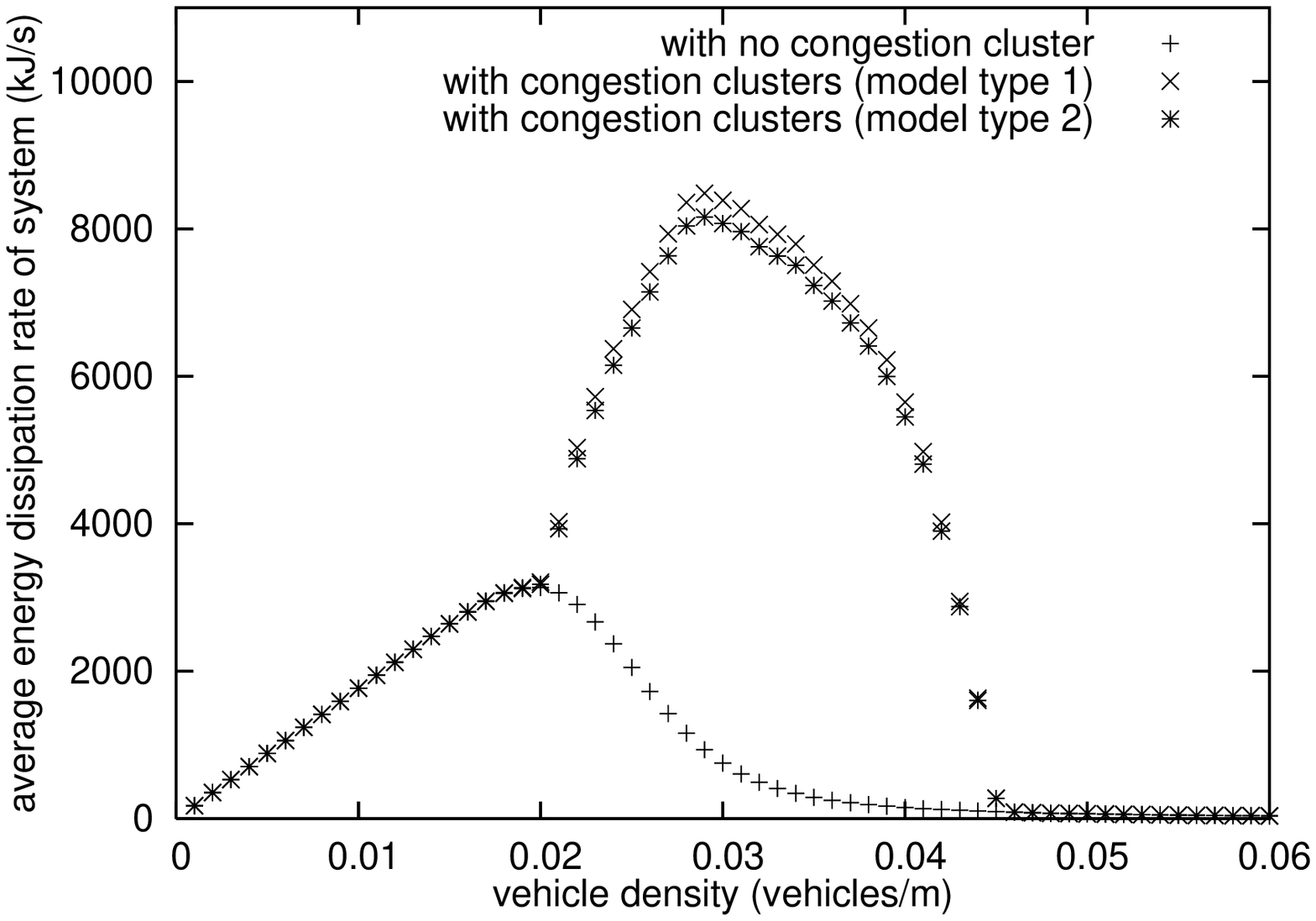}(b)\includegraphics[scale=0.4,clip,angle=-90]{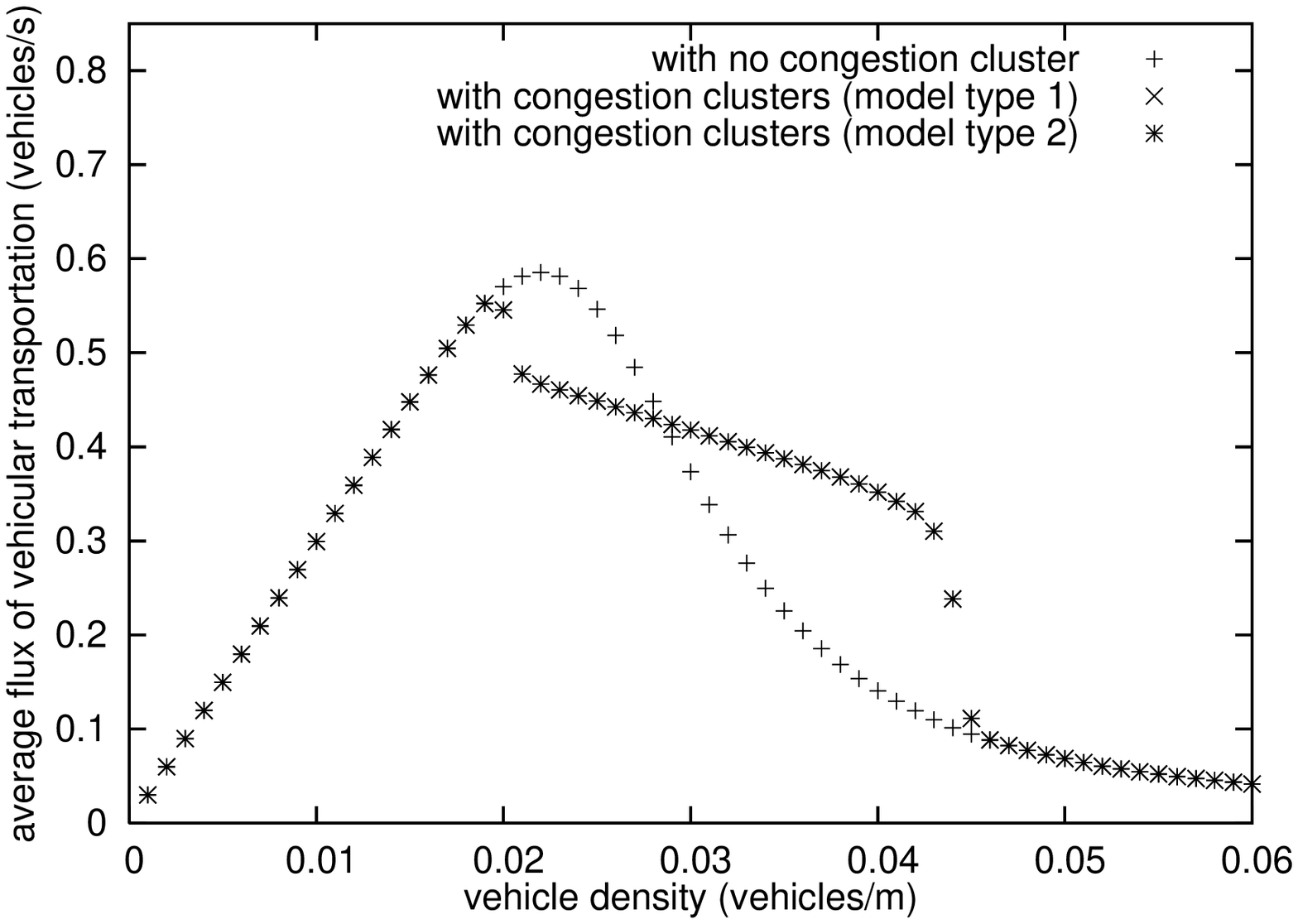}
\\
 \ \ 

(c)\includegraphics[scale=0.4,clip,angle=-90]{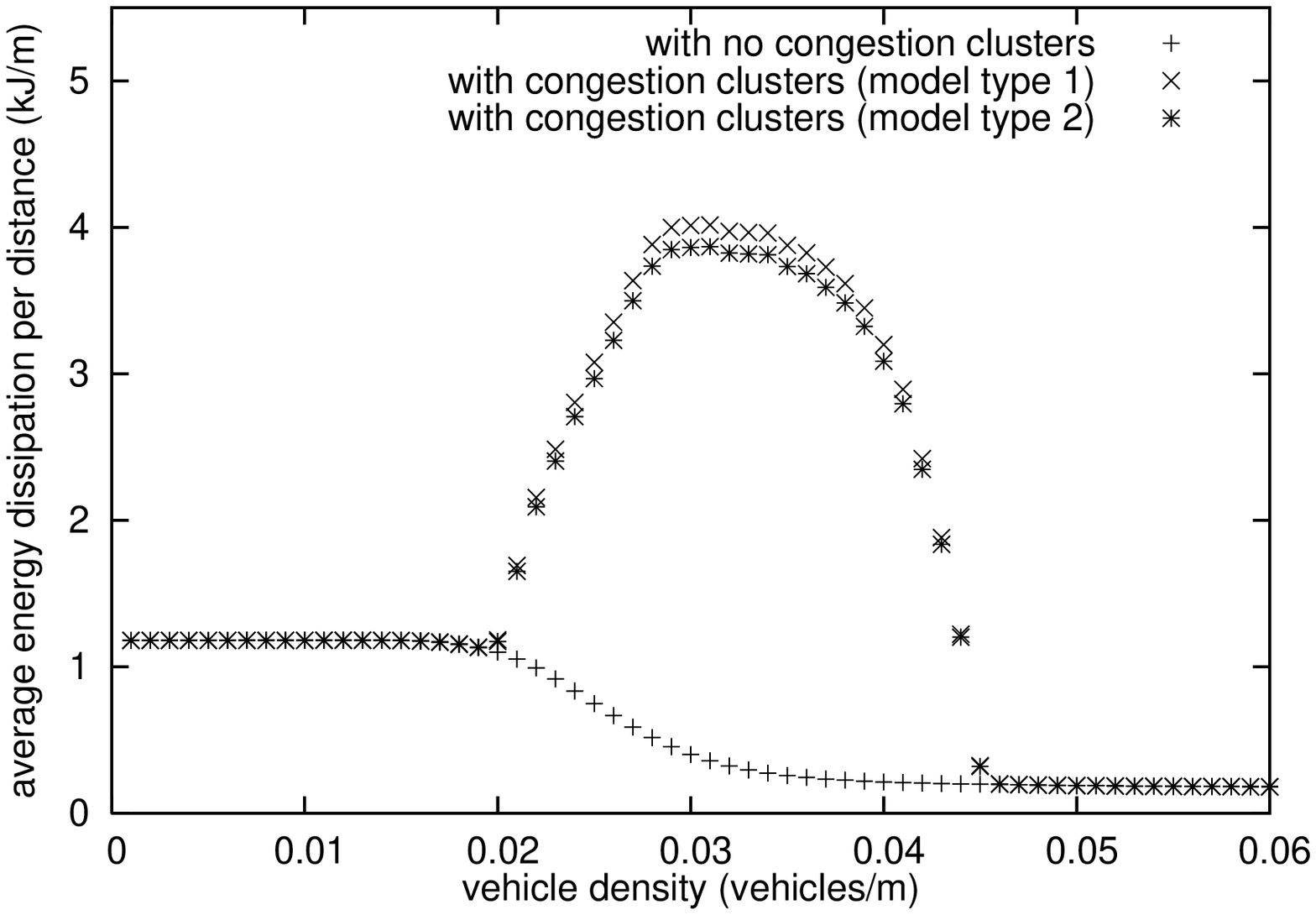}
\end{center}
\caption{Dependence of macroscopic values on the homogeneous vehicle density.   
(a) $\langle J_{q} \rangle$,  
(b) $Q$ and 
(c) $e$ 
vs $\rho$ respectively. 
Each figure contains plots of with (the case of $a=1$ s$^{-1}$) and without (the case of $a=5$ s$^{-1}$) congestion. 
Two kinds of plots having a congestion region correspond to two types of braking force model given in eqs. (14) and (15). }
\end{figure}

\begin{figure}[htbp]
\begin{center}
(a)\includegraphics[scale=0.40,trim=0 20 30 0,angle=-90]{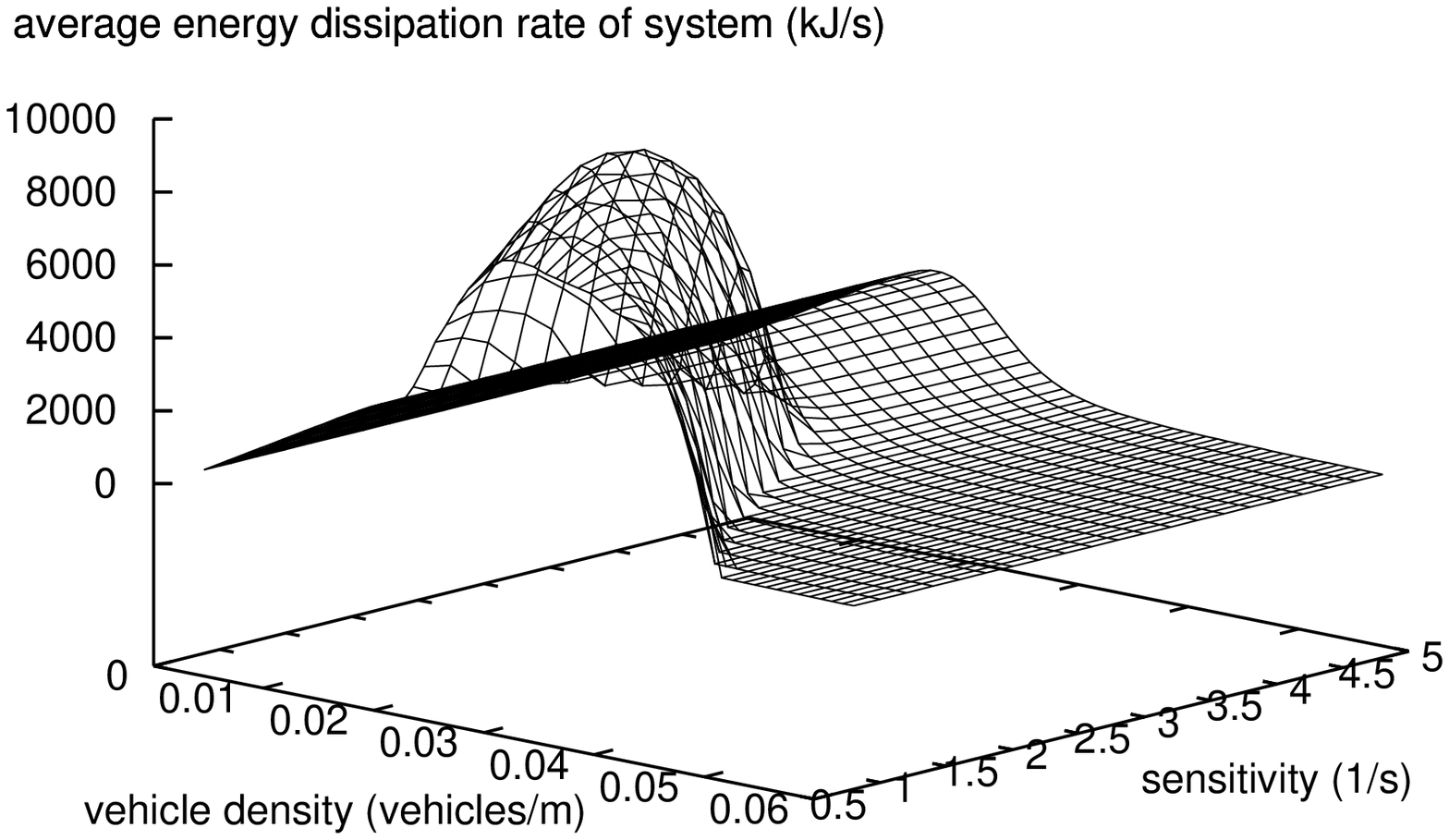}(b)\includegraphics[scale=0.40,trim=0 30 0 0,angle=-90]{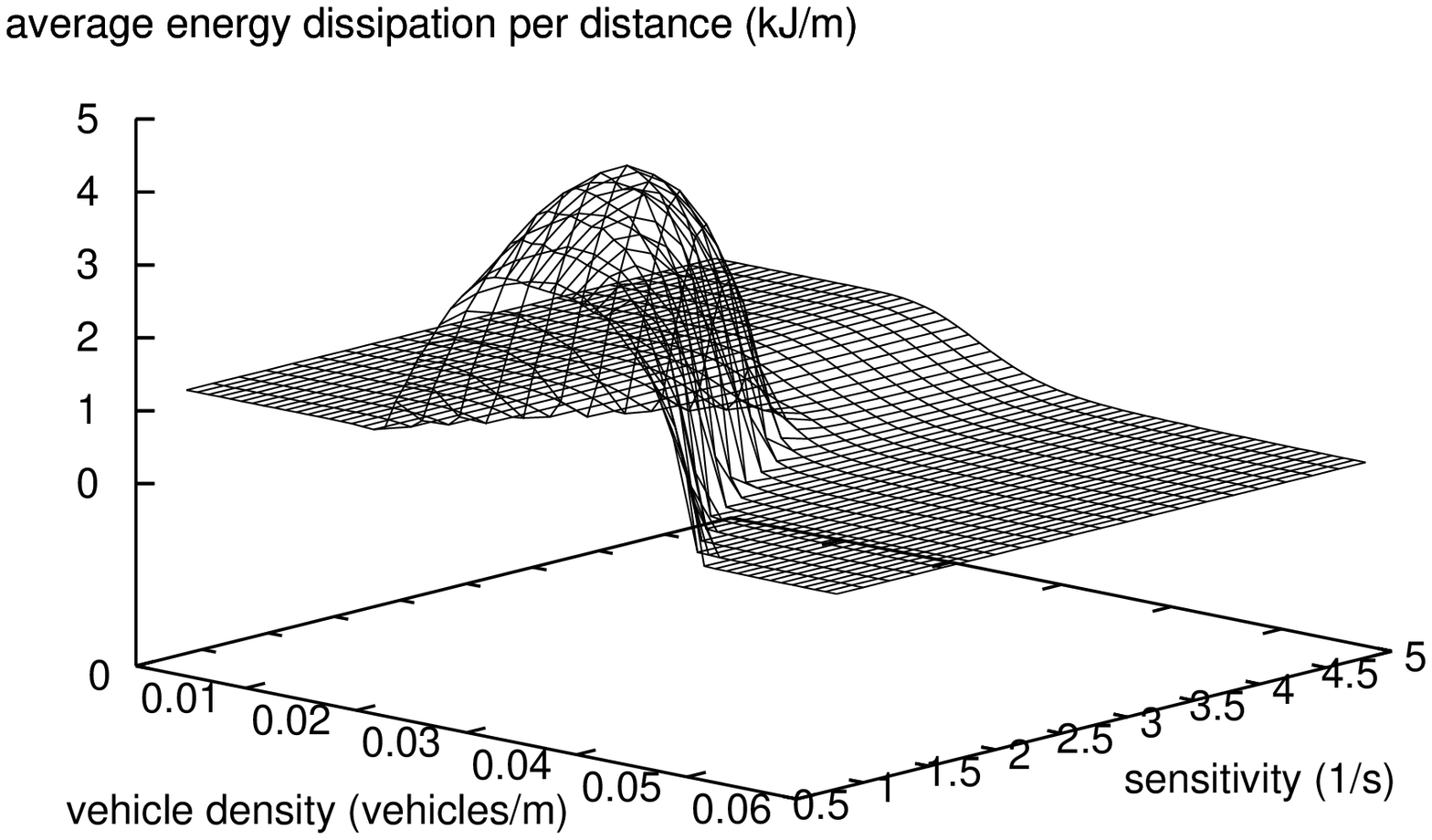}
\end{center}
\caption{Parameter dependence of 
(a) the average energy dissipation rate of system $\langle J_{q} \rangle$, and
(b) the average energy dissipation per distance $e$ 
in the space of $\rho \in [0,0.06]$ (the homogeneous vehicle density in circuit) and $a \in [0.8,5.0]$ (sensitivity: representing response speed of the driver). }
\end{figure}

\begin{figure}[htbp]
\begin{center}
(a)\includegraphics[scale=0.5,clip,angle=-90]{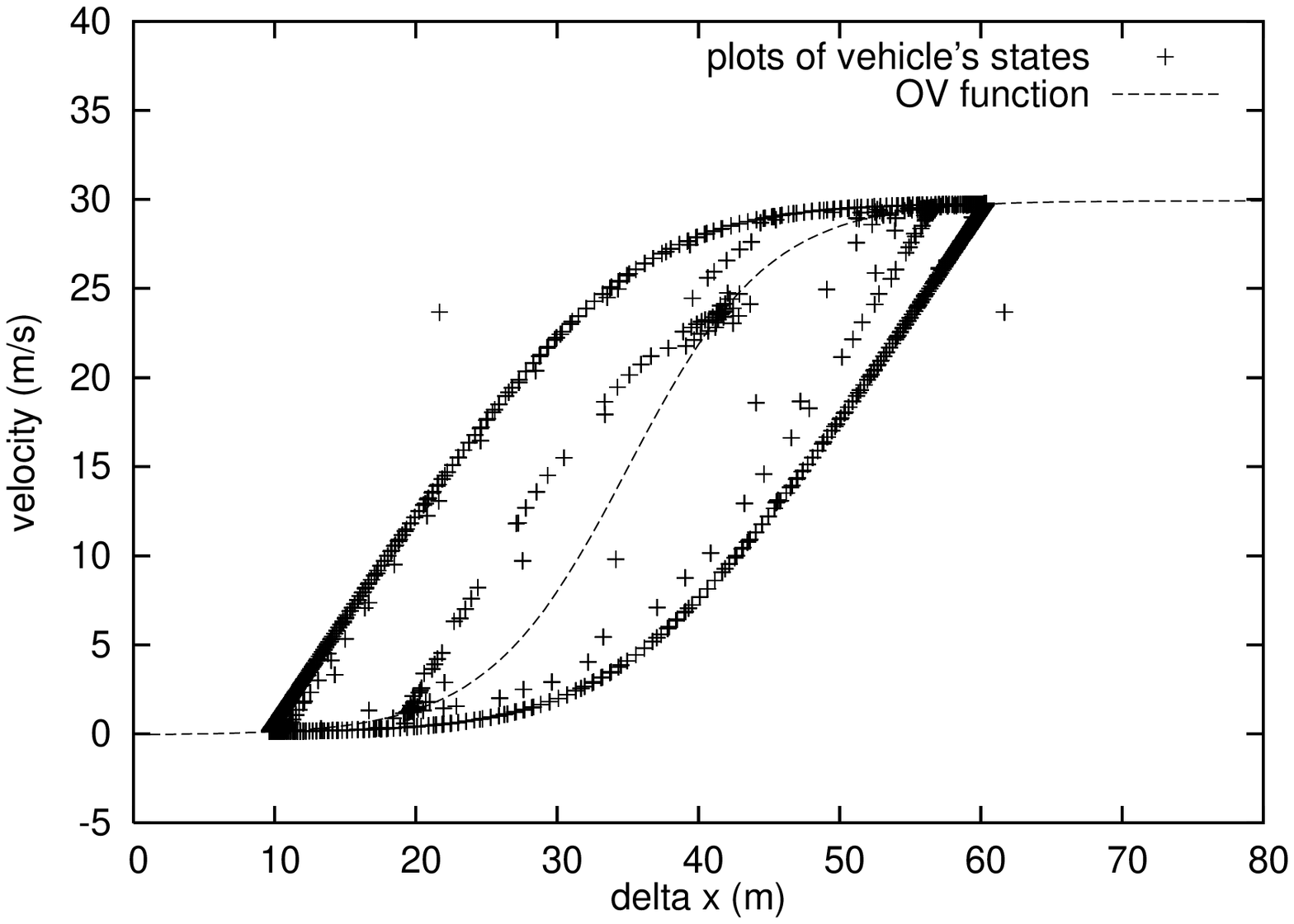}
(b)\includegraphics[scale=0.55,clip,angle=-90]{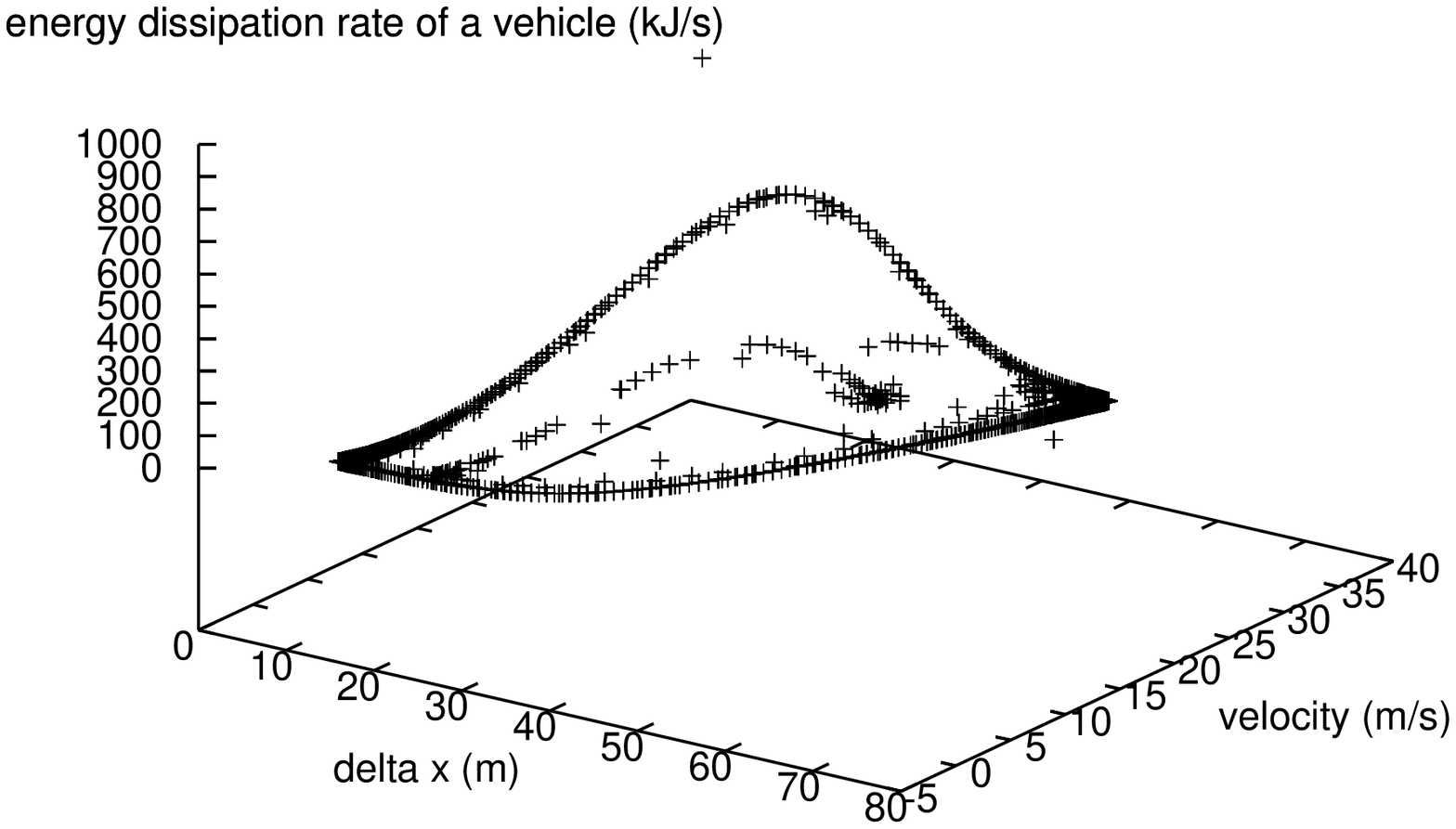}
\end{center}
\caption{Plots of the vehicle state in the phase space corresponding Fig. 2(c):  
(a) velocity of a vehicle $v$ vs forward distance $\Delta x$, 
and (b) energy dissipation rate of a vehicle $j_{q}$ as a function of $v$ and $\Delta x$.}
\end{figure}

\pagebreak

\section{Discussion}
We first discuss the implication of Fig. 2.
According to Bando et al.\cite{3,4}$^{)}$, the OV function and sensitivity $a$ affect whether congestion clusters appear or not. 
On the other hand, initial conditions of vehicles affect how congestion appears, in particular the number of simultaneous congestion clusters. 
The energy dissipation rate for a vehicle shows a spike when it reduces the speed, 
and consequently, the dissipation rate for the whole system shows a burst on congestion. 
These effects mean that the energy dissipation by braking force is much more responsible for the total energy dissipation 
than by air drag and frictions despite on account of relatively short duration of appearing $F_{b}$ in eqs. (14) and (15). 
It owes to the fact that $F_{b}$ is larger in the order of magnitude 
compared to other resistant forces as shown in eqs. (22), (23) and (24). 
The same reason applies the fact that simulations with eq. (14) give almost the same results as that with eq. (15).  
Additionally, a second order term of velocity $v$ in the air drag force $F_{a}$ affects the increasing of energy dissipation 
because the average $F_{a}$ at the situation of congestion becomes higher than at the situation of steady state flow. 
According to Table I and above discussion, 
we can say that energy dissipation burst is the feature of vehicular system with congestion: 
the less congestion clusters appear, the less the total energy dissipation becomes. 

Although the results of simulations appearing in Table I and in Fig. 3 indicate that 
the flux of vehicles is not so different between at the state of congestion and the state of steady flow 
and relatively independent of the number of congestion clusters,
the energy dissipation rate becomes much higher at the state of congestion than the state of steady flow.
Fig. 5 shows all vehicle's behavior in the phase space of the forward distance and velocity $(\Delta x, v)$ with time development 
on the state of congestion. 
It shows a hysteresis loop around OV function. 
This closed curve has two cusp point: 
the upper one shows large $\Delta x$ and $v$ which means vehicles are outside of congestion 
and the lower one shows small $\Delta x$ and $v$ which means inside of congestion. 
When vehicles are entering or leaving a cluster of congestion, they move counterclockwise on this loop.  
According to ref. 4, vehicles always follow this loop as long as sensitivity $a$ and OV function $V(\Delta x)$ are unchanged. 
From this scenario and Fig. 5(b) we recognize that 
vehicles show almost the same behavior along this loop when they enter any congestion cluster 
and reveal almost the same magnitude of energy dissipation burst because of eqs. (9) and (11). 
We can say that the energy dissipation rate $\langle j_{q} \rangle$ and $\langle J_{q} \rangle$  
are in proportion to the incidence of congestion clusters
because they trace this loop faithfully whether the number of appearing clusters is much or less. 
It corresponds to the results of numerical calculation of energy dissipation in Table I. 
Fig. 6 additionally shows that the energy dissipation per distance per vehicle $e$ is also in proportion to number of clusters 
beside the average flux of vehicular transportation $Q$ keeps its value almost constant. 
Furthermore, Fig. 4 shows that congestion and energy dissipation burst do not appear if sensitivity $a$ of OV model is large enough.

\begin{figure}[tbp]
\begin{center}
(a)\includegraphics[scale=0.4,clip,angle=-90]{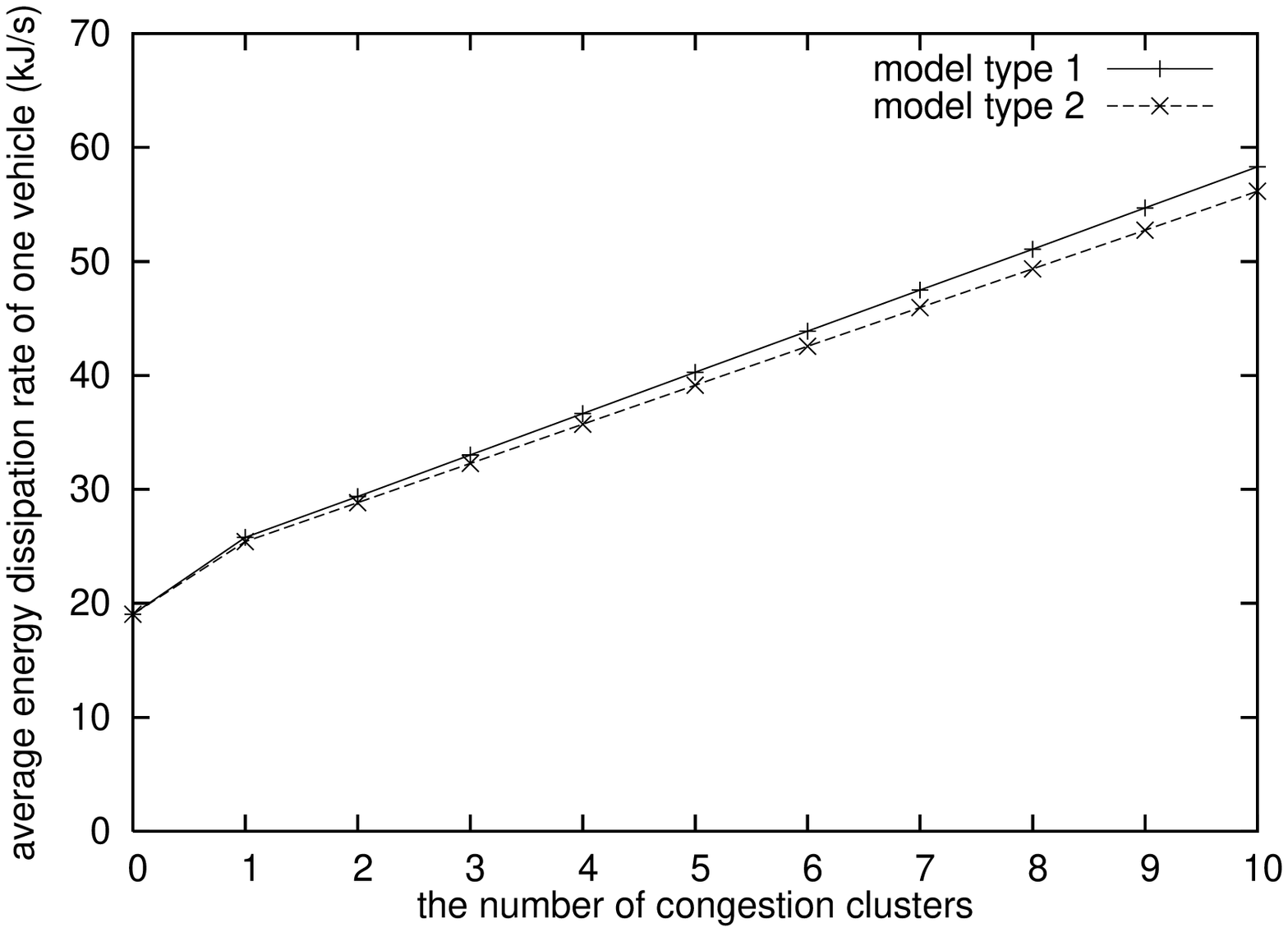}(b)\includegraphics[scale=0.4,clip,angle=-90]{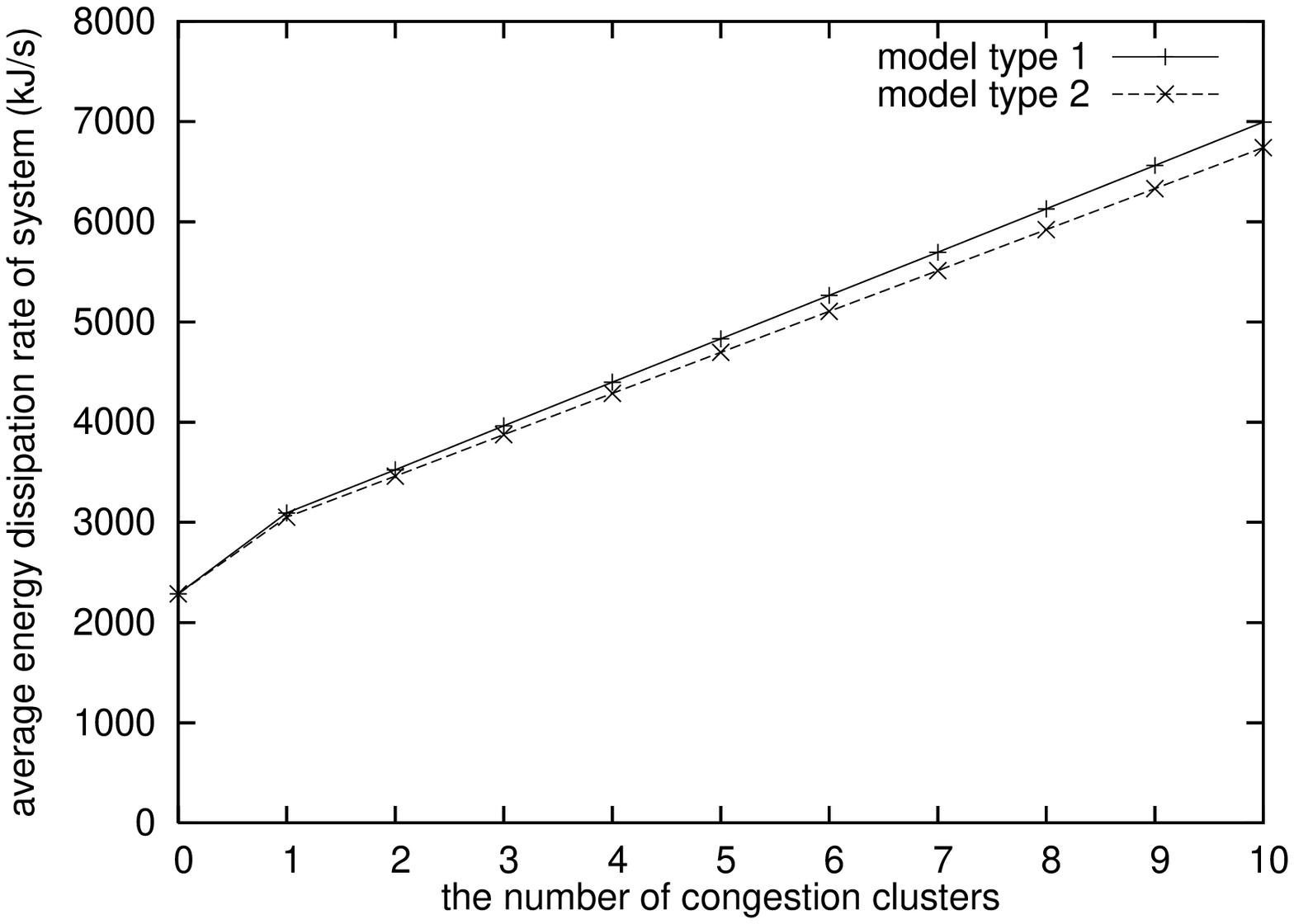}
(c)\includegraphics[scale=0.4,clip,angle=-90]{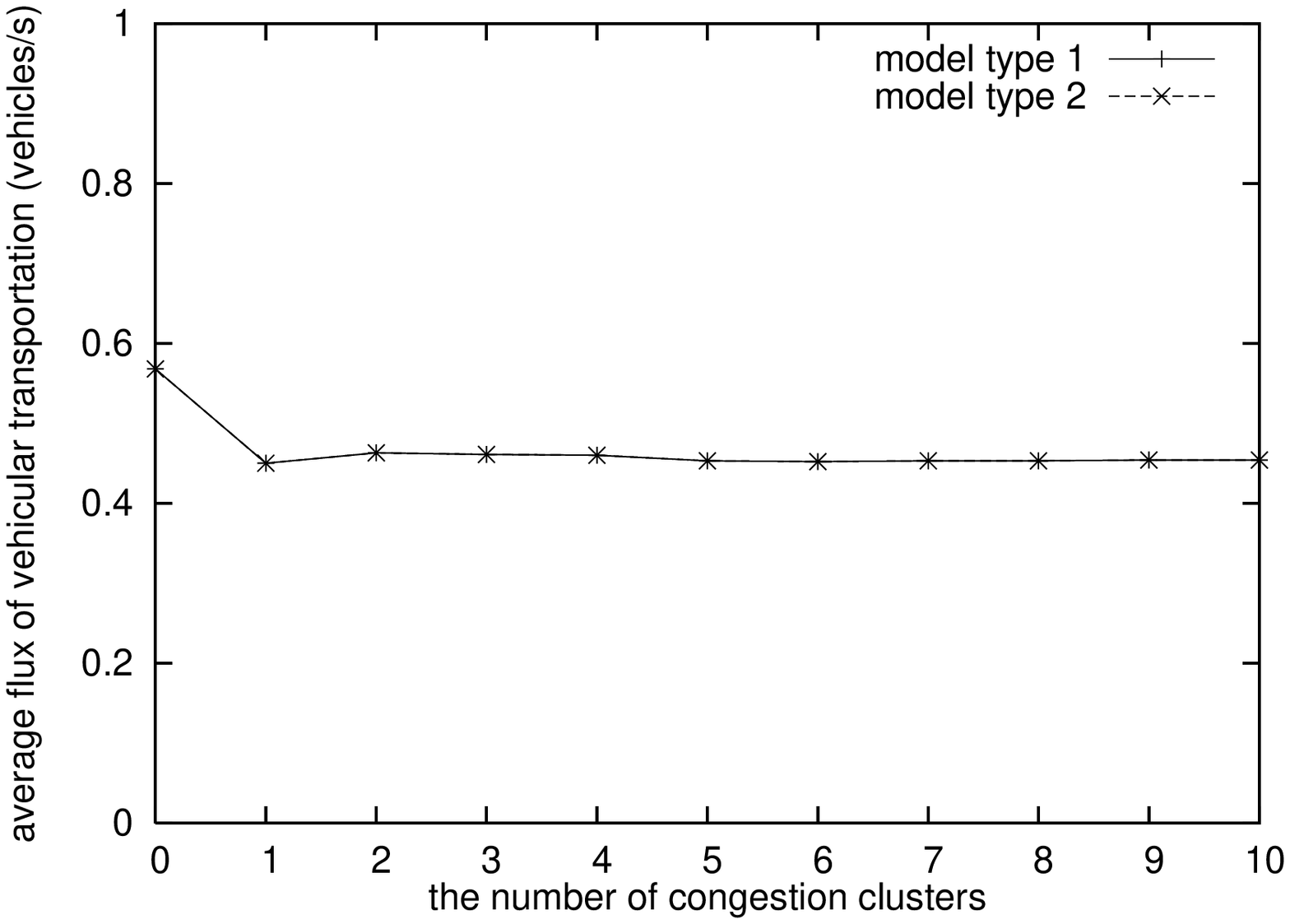}(d)\includegraphics[scale=0.4,clip,angle=-90]{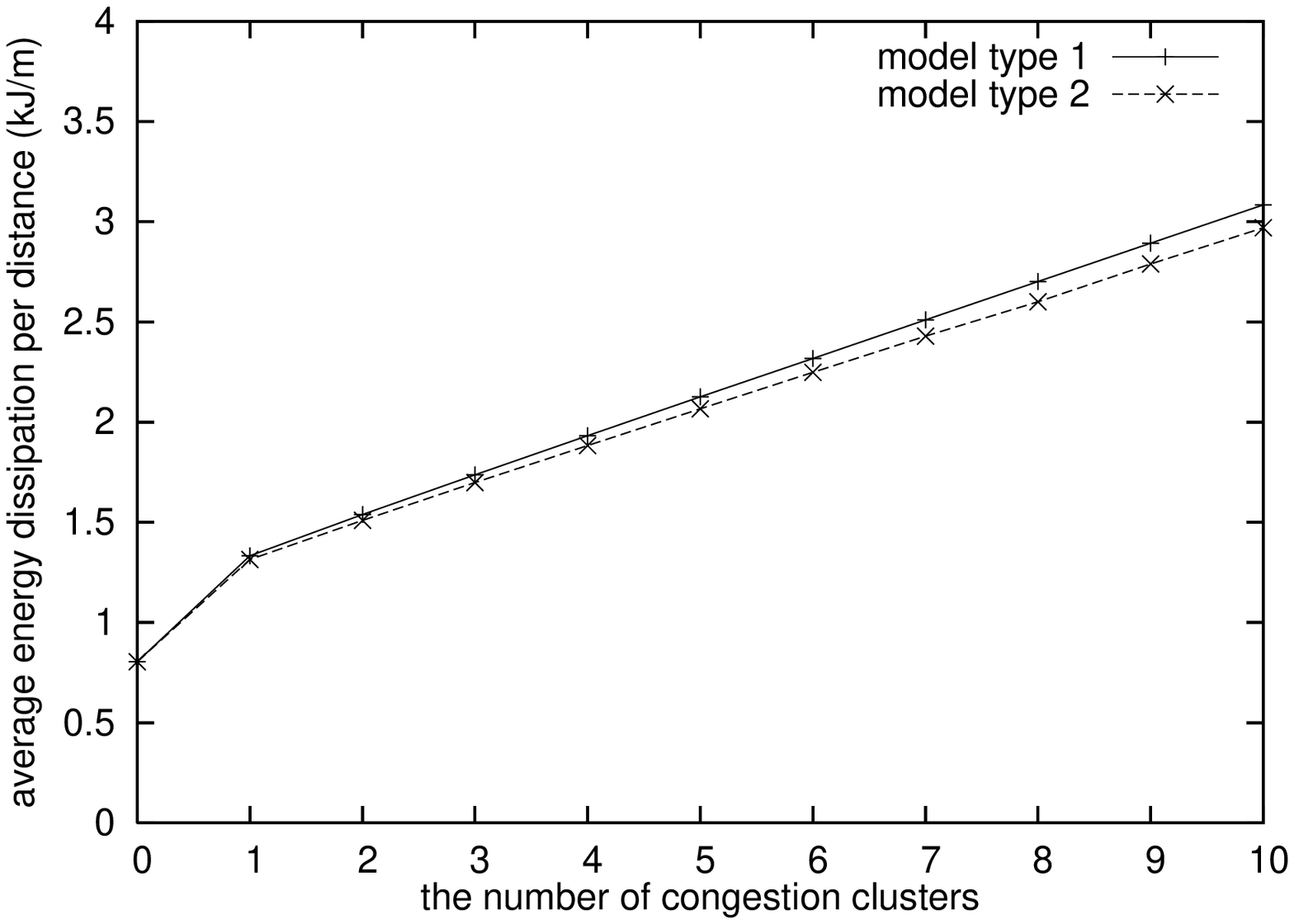}
\end{center}
\caption{Dependence of macroscopic values in Table. I:  
(a) $\langle j_{q} \rangle$, 
(b) $\langle J_{q} \rangle$, 
(c) $Q$ and 
(d) $e$
on the number of appearing congestion clusters in system. 
The figures show that $\langle j_{q} \rangle$, $\langle J_{q} \rangle$ and $e$ 
are in proportion to the number of appearing congestion clusters 
beside $Q$ keeps its value almost constant. }
\end{figure}

According to the above discussions, 
we may conclude that we can reduce the total energy dissipation of each vehicle and therefore of entire system in two ways. 
First, the sensitivity $a$ would be large enough so that no congestion appears nor the resulting energy dissipation burst. 
Second, when $a$ is not large enough, 
we can reduce energy dissipation if we can control the incidence of congestion clusters 
through controlling the initial positions of vehicles or by some other means. 
In other words, when the traffic is crowded in the real traffic expressway, 
you may be able to reduce energy waste without reducing transportation rate 
if you stop traffic flow at somewhere intentionally and make one big congestion.

But there still remains some questions. 
First, one might think that we should consider the thermal efficiency of the engine of a vehicle in our model 
because energy dissipation from the engine itself must occur when it works. 
In a real vehicle, the energy dissipation from engine may show higher rate than that considered in our model 
because the maximum thermal efficiency of normal vehicular gasoline engine reaches just 28-33\% \cite{10}$^{)}$.  
It will not bring an essential difficulty if the thermal efficiency shows constant value regardless of the state of vehicle: 
we can calculate entire energy dissipation from our simulation results by multiplying the inverse of thermal efficiency. 
However, it might cause a qualitatively different results if the thermal efficiency depends on the state of vehicle 
and our simulation results may be changed. 
One thing can be mentioned here that 
thermal efficiency has upper limit, which means energy dissipation can be increased by the state of vehicles 
but never decreased because of variation of thermal efficiency. 
Therefore the feature of energy dissipation burst will not be changed 
even if a model contains the variation of the thermal efficiency explicitly. 

There are also some problems on using models and simulation conditions. 
In this paper, 
we solve the OV model numerically with periodic boundary conditions and calculate the energy dissipation with a lot of parameters fixed. 
The sequence of moving vehicles is deterministic as long as we use the conventional OV model 
because it is described in the form of differential equations, 
while a real traffic may contain some kind of noise inside the system. 
This effect may suppress the burst 
since the proper inclusion of noise to the traffic system may prevent appearing of congestion 
and consequently the energy dissipation burst. 
Although the doubt is not dispelled completely, 
it can be said that the noise-induced acceleration and deceleration of vehicles would increase energy dissipation 
because of larger frequency of energy dissipation spikes.
Conversely speaking, we can reduce energy waste 
if we drive obediently under the rule of the OV model without time lags in the real world. 

There is also a problem of defining appropriate ensemble for averaging simulation conditions. 
The value and shape of diagrams in Fig. 4 might be changed 
if we put different ensemble though the feature of burst will not be changed. 
Therefore we have to consider how to set the ensemble of simulation condition 
to show whether the results of our simulation is universal in the OV model and applicable to the real world. 
There is also a remaining question why the right side of Fig. 2 shows fluctuations of the peak of $\langle j_{q} \rangle$, 
which may be due to the method of simulation but is not sure for now. 

For further studies, we should improve our simulation especially on how to control parameters and simulation conditions. 
Stochastically distributive parameters and open boundary conditions should be considered. 
We also have to improve our energy dissipation model in particular the modelling of resistant forces. 
It is also interesting to combine our energy dissipation model to some other models, 
for example the coupled map optimal velocity model (CMOV model) with random noise on velocity of vehicles \cite{7}$^{)}$.

\section{Summary}
We have presented an energy dissipation model for traffic flow based on the one dimensional optimal velocity model (OV model).
Being simple and well describing the appearance of congestion clusters in the system, 
the OV model is well suited model to introduce energy dissipation.  
In our model, the energy dissipation of the whole traffic system is calculated 
through modelling the resistant forces which work to each vehicle. 
We found that the energy dissipation rate of each vehicle shows spike due to deceleration when entering congestion, 
and its behavior in the phase space is almost always the same. 
Thus the energy dissipation spike is the characteristic property of one vehicle, 
and the energy dissipation of total system consequently shows burst when congestion appears. 
It is a characteristic feature of the whole vehicular system. 
The energy dissipation rate is in proportion to the incidence of congestion clusters.
This implies that we can reduce energy dissipation of the traffic system with preserving flux of vehicular transportation 
if we can control the number of traffic congestion clusters contained in entire traffic system.

\section{Acknowledgement}
The authors would like to thank 
Dr. Masako Bando for her comments about models and some realistic issues.

\end{document}